\documentclass[pre,twocolumn,aps,showpacs]{revtex4}
\usepackage{graphicx}
\usepackage[activeacute,english]{babel}
\begin{document}

\title {White noise flashing Brownian pump}
\author{
A. Gomez-Marin
 and J. M. Sancho}
\affiliation{
Facultat de Fisica, Universitat de Barcelona, Diagonal 647, 08028 Barcelona, Spain}

\date{\today}
     
\begin{abstract} 

A Brownian pump of particles powered by a stochastic flashing ratchet mechanism is studied. The pumping device is embedded in a finite region and bounded by particle reservoirs. In the steady state, we exactly calculate the spatial density profile, the concentration ratio between both reservoirs and the particle flux. A simple numerical scheme is presented allowing for the consistent  evaluation of all such observable quantities. 
\end{abstract}


\maketitle

\section{Introduction}
\label{sect:intro}  

Free diffusion by itself is not the appropriate physical mechanism for selective transport processes.
The appearance of net directed motion needs the 
breaking of detailed balance and spatial inversion symmetry. 
Inspired on Feynman's ratchet and pawl device \cite{feyn}, 
such phenomenon has been named after the ratchet effect. 
Moreover, when the environment is thermally fluctuating,  
Brownian transport is crucially affected and controlled by noise
(for a broad review see  \cite{reim} and references therein).
Its relevance to molecular engines, such as motors, pumps and channels, has been proposed from the foundation idea of the so-called Brownian motors \cite{oster}. 

\begin{figure}
\begin{center}
\begin{tabular}{c}
\includegraphics[scale=.30,angle=0]{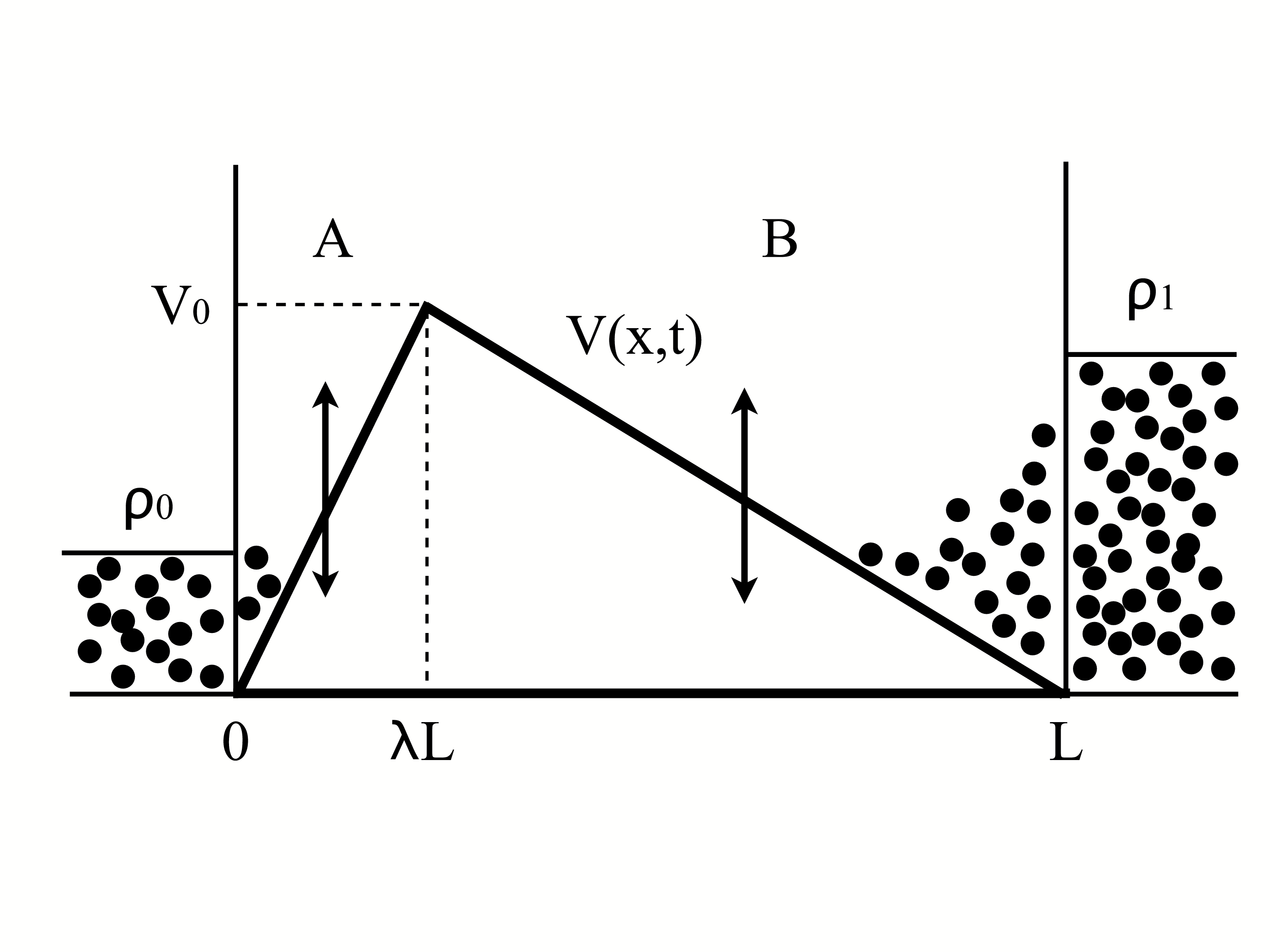}
\end{tabular}
   \end{center}
  \caption{Scheme of the pumping device: a spatially asymmetric and time dependent potential
 $V(x,t)$, embedded in a finite region of length $L$
 (denoted as the membrane) flashes in time 
creating a density profile between the reservoirs of densities $\rho_0$ and $\rho_1$.}
  \label{pot}
\end{figure}

Brownian pumping is then an active nonequilibrium transport process in which fluctuations play a very important role. 
Regarding experiments, Na,K--ATPase pumps have been perturbed by an oscillating electric field \cite{liu,tsong},
driving ions whose net flux was measured as a function of the amplitude and frequency of the field. 
The flux of particles created by pumping machines has been studied theoretically in \cite{pro,astum,astum0,kost}.
A typical ratchet mechanism is often assumed in the modeling and so  
normalization of the probability distribution and periodic boundary conditions are imposed.
In this work, the scope is different. While the ratchet mechanism is
still the responsible for the transport of particles, we focus on the nonequilibrium concentration gradient created and maintained by the pump at both sides of the membrane.
The system is considered finite and not infinitely periodic.
Then, our main observable is not only the flux $J$ of particles but the density profile $\rho(x)$ in the membrane and the ratio of concentrations $\rho_{1} / \rho_{0}$ created at the particle reservoirs. 
See figure \ref{pot}.

The structure of paper is the following. First we introduce the model for the Brownian pump by means of a Langevin equation, which can be mapped into a Seebeck ratchet \cite{buti1,buti2,sbm}. 
In the steady state, a standard theoretical analysis is carried out to obtain the non-equilibrium density of particles $\rho(x)$ that is generated at zero and finite flux. The flux $J$ itself is also determined.
Then we present a simple numerical scheme from which the densities and fluxes can be measured and satisfactorily compared to the  predictions.
For a piece--linear saw--tooth potential we calculate explicitly the expressions derived in the analysis and explore them as a function of the main parameters of the system. 
We end with some conclusions and comments for future work.

\section{General theoretical analysis}

We consider an underdamped Brownian particle moving under a time dependent potential $V(x,t)$. The corresponding equation of motion for its position is the Langevin equation
\begin{equation}
m\ddot{x}= -\gamma \dot{x}-V'(x,t)+\eta(t),
\end{equation}
where $m$ is the mass of the particle, $\gamma$ introduces the friction and 
$\eta(t)$ is the thermal noise accounting for thermal fluctuations  of the environment with the usual autocorrelation
\begin{equation}
<\eta(t)\eta(t')>=2\gamma k_{B}T \delta(t-t').
\end{equation}
The time dependent potential $V(x,t)$ consists of a ratchet part $V(x)$ (spatially asymmetric potential) time modulated by a stochastic  process in the following form, 
\begin{equation}
V(x,t) = V(x) [ 1 + \chi(t)],
\end{equation}
where $\chi(t)$ is another white and Gaussian noise with zero mean, uncorrelated with $\eta(t)$, whose autocorrelation is
\begin{equation}
<\chi(t)\chi(t')>=2Q \delta(t-t').
\end{equation}

In the regime in which friction dominates inertia,
the Langevin equation reduces to its overdamped limit,
\begin{equation}\label{under}
\label{main}
 \dot{x}=-V'(x)-V'(x) \chi(t)+\eta(t).
\end{equation}
Without  loss of generality, the friction constant $\gamma$ has been absorbed in the time units.
The last equation has a multiplicative noise and it should be treated with care. One must dilucidate first whether it has to be interpreted according to either Ito or Stratonovich rules.
The order of the limiting procedures from which one arrives to an overdamped equation from an underdamped equation, having assumed first that the noise $\chi(t)$ is obtained as the limit of an Orstein-Ulhenbeck process, determines that the appropriate stochastic interpretation is that of Ito \cite{sancho1,sancho2,pavlotis}.
Then, equation (\ref{under}) can be rewritten as
\begin{equation}
 \label{new}
{\dot x}=-V'(x)+g(x) \xi(t),
\end{equation}
where
\begin{equation}
g(x)=\sqrt{k_{B}T+Q\;V'(x)^{2}},
\end{equation}
and the new effective white noise $\xi(t)$ has zero mean and correlation $<\xi(t)\xi(t')>=2\delta(t-t')$.
The continuity equation for the density of particles $\rho(x,t)$  (the corresponding Fokker--Planck equation) is  
\begin{equation}
\partial_{t} \rho(x,t)=-\partial_{x}J(x,t),
\label{FP1}
\end{equation}
which, by using Ito's prescription in (\ref{new}), gives the following explicit expression for the flux $J(x,t)$ \cite{risken}
\begin{equation}
-J(x,t)= V'(x)\rho(x,t)+ 
\partial_{x} \left[ g(x)^2\rho(x,t)\right].
\end{equation}

In the steady state, the density is just a function of space and thus the flux becomes a constant, $J$. The density $\rho(x)$ follows a first order non homogeneous linear differential equation, whose formal solution is
\begin{equation}
\rho(x)=\mathcal{Z}(x,c_0,J)
\exp \left[ -\int_{0}^{x}\left(\frac{V'(z)}{g^{2}(z)}+ 2\frac{g'(z)}{g(z)}\right)dz \right]
\label{main2}
\end{equation}
with 
\begin{equation} \label{main2bis}
\mathcal{Z}(x,c_0,J)=c_0-J
\int_{0}^{x} \frac{dz}{g^{2}(z)}
 e^{\int_{0}^{z}\left(\frac{V'(z')}{g^{2}(z')}+2\frac{g'(z')}{g(z')}\right)dz'}.
\end{equation}
The unknown constant $c_0$ is found by imposing the left reservoir concentration $ \rho_0 \equiv \rho(0^-)$, as a fixed boundary condition.
Then $c_0=\rho_0$.
At each of the membrane boundaries with the reservoirs, this is at $x=0$ and $x=L$, we will distinguish between approaching from left and right side because of the possible existence of discontinuities, which we will see in short.
In what follows, we will study two different situations.

First, we impose a zero total flux: $J=0$. This corresponds to the case in which the pump in maintaining the maximum concentration difference between the two reservoirs across the membrane with no net leaking of particles. This situation is analogous the stalling force in Brownian motors.
From  (\ref{main2}), the density profile in the membrane is 
\begin{equation} \label{rox}
\rho(x)=\rho(x_0)\, \left[\frac{g(x_0)}{g(x)}\right]^2  \exp \left[ -\int_{x_0}^{x}\frac{V'(z)}{g^{2}(z)}dz \right].
\label{rhox}
\end{equation}
The exponent $2$ in the prefactor before the exponential is a characteristic of the Ito's interpretation. It changes to $1$ for Stratonovich's.  
One can find the expression for the ratio of concentrations at both sides of the pumping mechanism, being 
$\rho_1 \equiv \rho(L^+)$. 
Assuming no systematic drift in the system, this is $V'(L^+)=V'(0^-)$, then $g(L^+)=g(0^-)$ and therefore (\ref{rhox}) yields
\begin{equation}
\frac{\rho_1}{\rho_{0}}
=\exp \left[\frac{1}{k_BT}   \int_{0^-}^{L^+}\frac{-V'(z)}{1+ \frac{Q}{k_BT}V'(z)^{2}}dz\right],
\label{rho102}
\end{equation}
We will study such expressions in more detail in subsequent sections.

The concentration $\rho_0$ was fixed in the former case. If we now impose an arbitrary $\rho_1$ too, then the flux $J$ cannot longer be zero. From (\ref{main2}) and (\ref{main2bis}), one can get the explicit expression
\begin{eqnarray} \label{Jteoo}
J=\frac{\rho_{0}-\rho_{1}e^{
 \int_{0^-}^{L^+}\left(\frac{V'(z)}{g^{2}(z)}+2\frac{g'(z)}{g(z)}\right)dz   }} {
 \int_{0^-}^{L^+} \frac{1}{g^{2}(z)}
 e^{\int_{0^-}^{z}\left(\frac{V'(z')}{ g^{2}(z')}+2\frac{g'(z')}{g(z')}\right)dz'}dz   }.
\end{eqnarray}
Substituting $J$ back in (\ref{main2bis}) we can obtain, from (\ref{main2}), the density $\rho(x)$  in the steady profile for any choice of $\rho_0$ and $\rho_1$. 

Finally, let us stress that although the resolution of the Fokker--Planck equation is standard, the conditions we have imposed are not the usual ones in ratchet models, leading to new solutions.
The system is not infinitely periodic and the probability distribution function is viewed as a density of particles and thus, it is not normalized.  The present scheme is a realistic one for modeling pumps and channels operating between particle reservoirs.

\begin{figure}
\begin{center}
\begin{tabular}{c}
\includegraphics[scale=.30,angle=270,keepaspectratio=true]{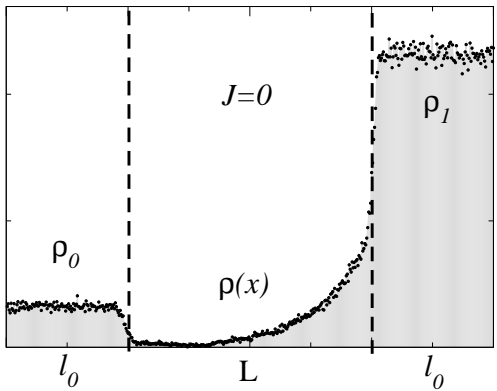}
\end{tabular}
   \end{center}
  \caption{Density profile $\rho(x)$ in the steady state at $J=0$
 obtained from numerical simulations. We use the potential introduced in Section IV. Reflecting boundary conditions are implemented at the end of both reservoirs.}
  \label{sims}
\end{figure}

\section{Numerical scheme}

The time discretization and simulation of the particle dynamic evolution equations, (\ref{main}) or (\ref{new}), are standard. 
The simplest method consists in using Euler's algorithm so that, for instance, from equation (\ref{new}), the evolution of the position $x_i(t) $ of the $i$--th particle turns out to be
\begin{equation}
x_i(t+ \delta t) = x_i(t) - V'(x_i(t)) \delta t + g(x_i(t)) X_i(t),
\label{algo}
\end{equation}
where $\delta t $ is the (small) integration time step and the stochastic term $X_i(t)$ is constructed as,
\begin{equation}
X_i(t)=\int_t^{t+\delta t} \xi(t') dt'= \sqrt{2 \delta t} \; \alpha_i, 
\end{equation}
in which $\alpha_i$ are gaussian random numbers $N(0,1)$. In fact, algorithm (\ref{algo}) corresponds to Ito's interpretation.

The implementation of the boundary conditions needs a more careful study. 
The membrane is surrounded by the reservoirs, which we also include virtually in the simulations, by means of an artificial length $l_0$. 
Then, the spatial domain of the system consists of three regions: the left reservoir $x\, \epsilon\, [-l_0,0)$, the membrane $x \, \epsilon \,(0,L)$ and the right reservoir $x \,\epsilon \,(L, L+l_0]$. 
We consider  a large number $N$ of non-interacting particles, each one evolving according to (\ref{algo}).
When a particle is in a reservoir then $V(x)=0$ and it diffuses freely; only random kicks due to thermal noise drive the particle.  
Each particle is numbered and its position, $x_i(t)$, recorded. 
The numerical simulation always starts with particles distributed homogeneously. The measures are taken waiting a certain time so that the system has reached the steady state.
In general we observe a transient during which $\rho_0(t)$ decreases and $\rho_1(t)$ increases up to the steady state, in which   
$\rho_1 > \rho_0 $ and it is kept constant. 
To obtain the density profiles $\rho(x)$ an histogram of the distribution of particles is drawn.

For the physical case of zero flux ($J=0$), we impose
reflecting boundary conditions at the points $x=-l_0$ and $x=L+l_0$ of our simulation system.
For non zero flux ($J\neq 0$), we take periodic boundary conditions at points $-l_0$ and $ L+l_0$, ensuring the same flux along all the regions. 
These two simple rules lead to the physical situations of interest, as we will show now.
In the former case (reflecting boundary conditions), the steady state density of particles at both reservoirs is constant, as expected for $J=0$. See Fig. \ref{sims}. 
One can plot the detailed shape of $\rho(x)$ in the membrane and also measure the concentration ratio $\rho_1/\rho_0$ that the pumping mechanism has created. We will analyze such observable quantities in detail in the following sections.
The difference of concentrations created and maintained implies that the flux was not zero during the transient and that the system is now  in a nonequilibrium steady state.

\begin{figure}
\begin{center}
\begin{tabular}{c}
\includegraphics[scale=.30,angle=270,keepaspectratio=true]{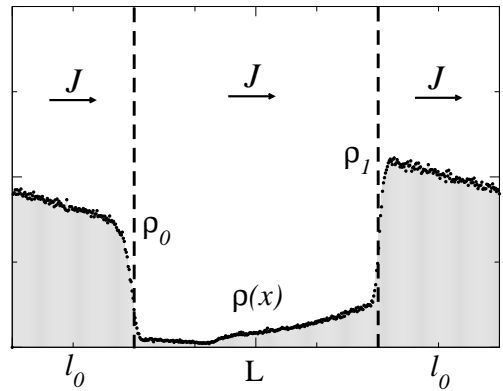}
\end{tabular}
   \end{center}
  \caption{Density profile $\rho(x)$ in the steady state at $J\neq0$
 obtained from numerical simulations as in previous figure. Now periodic boundary conditions are applied at the ends of the reservoirs   leading naturally to a constant $J$ and fixed $\rho_1$  and $\rho_0$.}
  \label{flux}
\end{figure}

When imposing periodic boundary conditions in the simulations (not in the real physical Brownian pump since $\rho_0\neq \rho_1$) we observe, in the steady state, a linear density profile in the reservoirs,  signature of the finite flux $J$.
Such flux should be constant and equal everywhere. In Fig. \ref{flux} a typical histogram is shown. 
This simple and intuitive method is able to generate data of $J$ as a function of the boundary conditions; the concentrations $\rho_0$ and $\rho_1$. More advanced and sophisticated studies of simulation of Langevin trajectories with specific boundary conditions can be found in Refs.\cite{sims1,sims2,sims3}.
In our method, the pump is moving particles from the left to the right forcing a gradient of concentration in the reservoirs (which are  connected only in the numerical scheme). Accordingly, a net flux appears, which fulfills Fick's law so that
\begin{equation}
J=\frac{k_B T}{2 l_0} \left( \rho_1 - \rho_0 \right).
\label{Fick}
\end{equation}

Therefore, independently of the initial conditions the system will evolve to the steady state in which the value of $J$ in equations (\ref{Jteoo}) and (\ref{Fick}) coincides.
Then the flux can be measured by counting the net number of particles crossing the point $x=L+l_0$ (where the periodic condition is implemented) as a function of time. A complementary way is to fit the  histograms in the region $(L,L+l_0) \bigcup (-l_0,0)$ with the linearly decreasing behavior predicted by Fick's law. From the slope, $J$ is found and also the concentrations $\rho_0$ and $\rho_1$ can be determined.
The agreement between theory prediction for $J(\rho_0,\rho_1)$ and simulation results is good as we will show in the next section.

\section{Analytical and simulation results}

In this section we complete the analytical results of Ref. \cite{SPIE} and compare them with numerical simulations. The explicit model we consider is a piece-linear saw-tooth potential depicted in Fig. \ref{pot}. It is defined in two regions as
\begin{eqnarray} \label{A}
V_{A}(x)=V_{0} \frac{x}{\lambda L} &   & x \; \epsilon \; (0,\lambda L),
\end{eqnarray}
\begin{eqnarray} \label{B}
V_{B}(x)=V_{0} \frac{L-x}{(1-\lambda)L } &  & x \; \epsilon \;
(\lambda L, L).
\end{eqnarray}
$V_{0}$ is the height of the potential, $\lambda$  (which can only take values between $0$ and $1$) controls the asymmetry and $L$ is the total length where the pumping device is allocated.

The forthcoming subsections are devoted to calculate and discuss the exact analytical expressions of the profile density of particles $\rho (x)$ and the  concentration ratio $\rho_1/\rho_0$  at $J=0$, as well as the normalized flux $J/\rho_0$ as a function of $\rho_1/\rho_0$.  Since the potential is linear in pieces, the force is discontinuous, yielding to discontinuities in the density profiles, which we will carefully characterize without any further problems. 

\subsection{Density profile $\rho(x)$ at $J=0$}

\begin{figure}
\begin{center}
\begin{tabular}{c}
\includegraphics[scale=.30,angle=270,keepaspectratio=true]{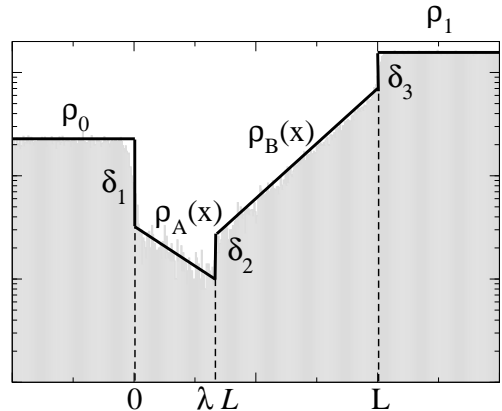}
\end{tabular}
   \end{center}
  \caption{Density profile $\rho (x)$ for $v_{0}=8$, $\lambda=1/3$ and 
 $\alpha=0.01$ in logarithmic scale. The histogram from simulations 
(in gray) falls just on top of the theoretical predictions (solid lines).}
  \label{regions}
\end{figure}

Let us recall expression (\ref{rhox}) for the spatial density profile $\rho(x)$ given a baseline concentration $\rho_0$ and zero flux conditions, $J=0$. From the potential in (\ref{A}) and (\ref{B}), $\rho(x)$ can be obtained exactly. Nevertheless as the potential is linear in pieces, we expect discontinuities in  $\rho(x)$. Thus, we have to evaluate the profile in four different zones denoted by the subscripts $0$, $A$, $B$ and $1$, whose meaning is clear from Figs. \ref{pot} and \ref{regions}.
First we introduce the following dimensionless parameters to simplify calculations,
\begin{eqnarray}
v_{0} \equiv \frac{V_{0}}{k_{B}T}, & &
\alpha \equiv \frac{Qk_{B}T}{L^{2}},
\label{voalpha}
\end{eqnarray}
where $v_0$ is the relative energy barrier of the
potential compared to thermal energy. The parameter  $\alpha$ is measure of the strength of the flashing mechanism.

The density of particles in region $A$, this is $\rho_A(x)$, in which $x \; \epsilon \; (0^+,\lambda L^-)$, is obtained from (\ref{rhox}) by taking $x_0=0^-$, so that $\rho(x_0)=\rho_0$ and $g(x_0)= \sqrt{k_B T}$. Noticing that $g(x)$ in region $A$ (denoted by $g_A(x)$) is a constant, one has
\begin{equation}
g_A=g(0^+) = g(\lambda L^-)=\sqrt{k_BT+Q\left( V_0/\lambda L\right)^2}
\end{equation}
which finally yields to
\begin{equation}
\rho_{A}(x) = \frac{\rho_0}{\omega_1 }
  \; \exp\left( - \frac{v_{0}}{\omega_1 }
  \frac{x}{\lambda L}   \right),
\label{solA}
\end{equation}
where, for simplicity in the notation, the following new dimensionless quantities have been defined,
\begin{eqnarray}
\omega_{1}\equiv 1+\alpha \left( \frac{v_{0}}{\lambda}\right)^2, & & \omega_{2}\equiv 1+\alpha \left( \frac{v_{0}}{1-\lambda}\right) ^{2}.
\end{eqnarray}
Then, the jump of the concentration at $x=0$ is simply
\begin{equation}\label{d1}
\delta_{1}\equiv \rho(0^{-})- \rho(0^{+})= \rho_{0} \left(1-
  \frac{1}{\omega_1 }      \right).
\end{equation}

The density of particles in region $B$ is found similarly. We recall (\ref{rhox}) and now choose  $x_0=\lambda L^-$. From the above expressions and by 
noticing again that $g_B(x)$ is a constant,
\begin{eqnarray}
g_B = \sqrt{k_BT+Q\left( V_0/(1-\lambda) L\right)^2},
\end{eqnarray}
the concentration profile $\rho_B(x)$ is found to be
\begin{equation} \label{solB}
\rho_{B}(x) = \frac{\rho_{0}}{\omega_2} \; \exp \left( - \frac{v_{0}}{\omega_1}  \right)  \;  \exp \left(   \frac{v_{0}}{\omega_2}
  \frac{x-\lambda L}{L-\lambda L}    \right).
\end{equation}
The jump between zones $A$ and $B$ (at $x=\lambda L$) is
\begin{eqnarray} \label{d2}
\delta_{2}\equiv \rho(\lambda L^{+})-\rho(\lambda L^{-})
= \rho_{B}(\lambda L) -   \rho_{A}(\lambda L)
\nonumber \\
 = \rho_{0} \; \exp  \left( - \frac{v_{0}}{\omega_1}  \right)  \;\left(
  \frac{1}{\omega_2 }  -  \frac{1}{\omega_1}  \right).
\end{eqnarray}

The constant density $\rho_1$ at the other side of the membrane is discussed in detail in the next section. The jump $\delta_3$, which corresponds to $x=L$, is obtained likewise from the difference between $\rho_1$ and $\rho_B(L)$. 
In Fig. \ref{regions} we show, in logarithmic scale, the above analytical predictions for $\rho(x)$ calculated in every region and the corresponding jumps. Note that every piece of prediction fits perfectly to the histogram (in gray) built from the data.

\subsection{Ratio $\rho_{1}/\rho_{0}$ at $J=0$.}

We focus now on the value of the ratio of concentrations at both ends of the membrane that the pumping Brownian device is able to create and maintain at the stalling regime.
For the linear saw-tooth potential,  equation (\ref{rho102}) gives
\begin{equation} \label{ratio}
\frac{\rho_{1}}{\rho_{0}}=\exp \left[ v_{0} \left(
\frac{1}{1+\alpha \left( \frac{v_{0}}{1-\lambda} \right)^{2} } -\frac{1}{
  1+\alpha \left( \frac{v_{0}}{\lambda} \right)^{2}  }  \right) \right].
\end{equation}

\begin{figure} [t]
\begin{center}
\begin{tabular}{c}
\includegraphics[scale=.30,angle=270,keepaspectratio=true]{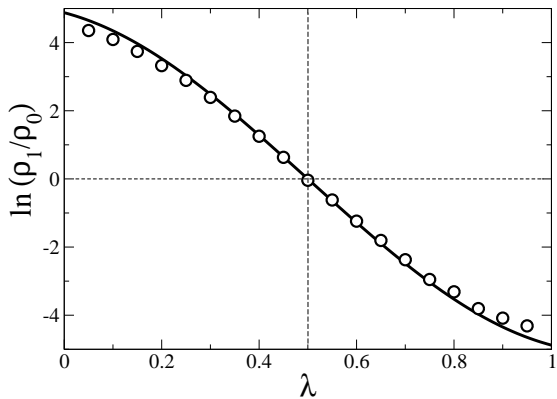}
\end{tabular}
   \end{center}
 \caption{Logarithm of the ratio of concentrations versus the
    asymmetry parameter $\lambda$ at $J=0$. 
Note that it is antisymmetric under the transformation $\lambda \leftrightarrow  1-\lambda$. The parameters are $v_{0}=8$ and $\alpha=0.01$. Solid line corresponds to theory, equation (\ref{ratio}), and circles to simulation data.
    }
  \label{lambda}
\end{figure}

Let us explore this result  on the parameters $v_{0}$, $\alpha$ and $\lambda$.
In Fig. \ref{lambda}  we check the symmetry properties of our prediction with respect the parameter $\lambda$. This figure shows the right-left inversion symmetry of the
problem when we change $\lambda$ for $1-\lambda$. Note that at $\lambda=0.5$,
although there is a time modulation of the potential, the device does not pump
because the spatial inversion symmetry is not broken and so there is no preferred direction. 
The more asymmetric the potential is, the greater the pumping capacity is achieved. This does not mean higher efficiencies with respect to energy consumption. This issue is not studied in this work.

\begin{figure} [t]
\begin{center}
\begin{tabular}{c}
\includegraphics[scale=.30,angle=270,keepaspectratio=true]{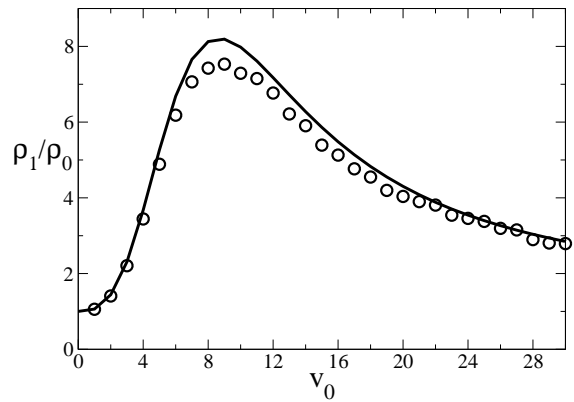}
\end{tabular}
   \end{center}
  \caption{Ratio of concentrations $\rho_1/\rho_0$ as a function of
    the relative potential barrier $v_0$ at $J=0$. The values of the parameters are $\lambda=1/3$ and $\alpha=0.01$. Line and circles as in previous figure. }
  \label{vo}
\end{figure}

In Fig. \ref{vo} the density ratio is studied versus the relative energetic barrier $v_0$. There is an optimum value which gives the largest difference. If the barrier height is small with respect the thermal energy, diffusive loses through the pump are important and   the ratio decreases. Note that for $v_0 \to 0$ there is no concentration difference; $\rho_1=\rho_0$. On the other hand, for very large values of $v_0$, few particles can cross the barrier and get to the other side despite the flashing of the potential. Then the pumping decreases again.

\begin{figure} [t]
\begin{center}
\begin{tabular}{c}
\includegraphics[scale=.30,angle=270,keepaspectratio=true]{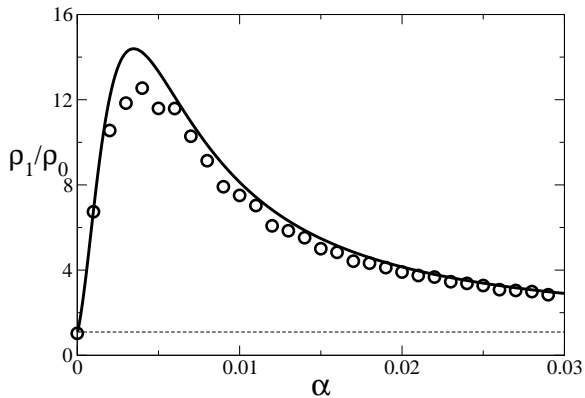}
\end{tabular}
   \end{center}
  \caption{Ratio of concentrations versus parameter $\alpha$ at $J=0$ for $v_{0}=8$ and $\lambda=1/3$. Line and circles as in previous figure.}
  \label{beta}
\end{figure}

In Fig.  \ref{beta} we show the ratio of concentrations versus $\alpha$  (we have varied $Q$ in the $\alpha$ exploration).
For low values of $Q$, the potential barely changes in time, diffusion dominates and thus particles can scarcely be pumped. 
For a very strong flashing the potential is so often distorted that particles do not have time to cross through the membrane and they are again poorly pumped.
In between both regimes, there is a region that enhances transport. Such optimal regime indicates that the flashing intensity $Q$ can be tuned to be optimal. 
This is a common feature of flashing ratchets \cite{reim}.
The maximum appeared in Figs. \ref{vo} and \ref{beta} can be compared qualitatively with the experimental results of \cite{liu,tsong} for the amplitude and frequency of the flashing perturbation.

Let us analyze in more detail the case $\alpha=0$, which gives $\rho_0=\rho_1$. This limit is physically interesting because it corresponds to $Qk_{B}T/ L^{2} \to 0$. When
the intensity in the multiplicative noise vanishes ($Q=0$), the breaking of detailed balance does not occur and, so, the ratchet effect cannot take place. 
Moreover, another way to make $\alpha$ vanish is setting $T=0$. We have to be very careful because $T$
also appears in $v_{0}=V_{0}/k_{B}T$.
In fact, in the absence of thermal fluctuations, the flashing ratchet mechanism still works because the multiplicative noise does all the job (breaks detailed balance and supplies fluctuations).  
Therefore, in the $\alpha$ exploration we have kept $T$ different from zero.

\subsection{ Flux $J/\rho_0$ versus  ratio $\rho_1 / \rho_0$ }

We can recall expression (\ref{Jteoo}) for the total flux of particles and rewrite it as 
\begin{eqnarray} \label{Jone}
\frac{J}{\rho_0} = \frac{1-\rho_1/ \rho_1^{stall}}{ \frac{1}{k_B T}\int_{0^-}^{L^+} 
 e^{\int_{0^-}^{z} \frac{V'(z')}{ g^{2}(z')}  dz'}dz},
\end{eqnarray}
where $\rho_1^{stall}$ is the concentration in the right reservoir  when $J=0$, which is calculated in (\ref{rho102}). 
For the linear saw--tooth potential, the above formal expression can be explicitly expressed as 
\begin{equation} \label{newJ}
J=\left( \frac{k_BT}{L} \right) \frac{\rho_0 e^{-v_0/\omega_1}-\rho_1e^{-v_0/\omega_2}}{\mathcal{N}_1+\mathcal{N}_2},
 \end{equation}
where
\begin{eqnarray}
\mathcal{N}_1=\frac{\omega_1}{v_0} \lambda \left(1-e^{-v_0/\omega_1}\right) ,
 \end{eqnarray}
\begin{eqnarray}
  \mathcal{N}_2= \frac{\omega_2}{v_0}(1-\lambda)
 \left( 1-e^{-v_0/\omega_2}\right) .
 \end{eqnarray}
Expression (\ref{newJ}) fulfills the symmetry $J\to -J$ when $\lambda \leftrightarrow 1-\lambda$ and $\rho_0 \leftrightarrow  \rho_ 1$, which means that if we take the mirror image of the set up, we should see the same flux going to the opposite direction.
Note that when $J=0$ only the ratio of densities is relevant, while for $J\neq 0$, both values are needed separately. 
There is a decreasing linear behavior of $J$ with respect to $\rho_1$, as it is clear from simple inspection of (\ref{Jone}) and (\ref{newJ}).
If $\rho_1 >  \rho_1^{stall}$  the flux is reversed because entropic forces surmount the pumping driving. 

\begin{figure} [t]
\begin{center}
\includegraphics[scale=.30,angle=270,keepaspectratio=true]{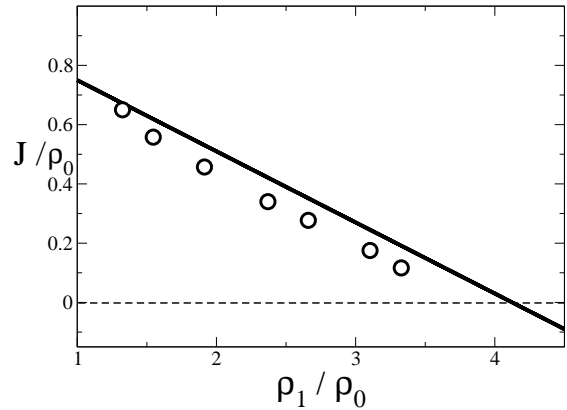}
   \end{center}
  \caption{Flux $J/\rho_0$ as a function of $\rho_1/\rho_0$. 
 The system parameters are  $\lambda=1/3$, $v_0=8$, $\alpha=0.02$ and $L=1$.  
 Solid line corresponds to theory, equation (\ref{newJ}), and circles to simulation data
}
  \label{Jrho}
\end{figure}

In Fig. \ref{Jrho} the normalized flux $J/\rho_0$ is plotted as a function of the density ratio $\rho_1/\rho_0$. 
The simulation points are successfully obtained from the method explained in the numerical scheme section. Increasing the length $l_0$ of the left and right reservoirs the value of the flux $J$ decreases. We can measure it from the slope of the linear profile of the density in such regions and also extract $\rho_0$ and $\rho_1$. Despite some small deviations due to the errors in the process of data obtention because of finite statistics and non zero $\delta t$, we can say that theory and simulations are in agreement. This confirms the validity of the numerical scheme proposed, which allows to measure fluxes and concentrations from simple numerical simulations which have a clear physical interpretation.

\section{Conclusions}

Brownian pumps, unlike typical Brownian motors, do not aim to create 
a net flux of particles in periodic boundary conditions, but instead to achieve and actively keep a density gradient between two reservoirs.
We have studied a simple model in which the ratchet effect together with appropriate boundary conditions leads to such mechanism.
It has been thoroughly characterized from the theoretical point of view with analytical exact results. Moreover, a new and simple numerical scheme has been proposed to measure concentrations and fluxes, faithfully reproducing all the theoretical predictions.
The efficiency of such devices is not easy  to investigate since one should analyze how much energy the fluctuating potential is inserting into de pump and what is the energetic  profit taken out from such input. This issue together with the use of physical parameters in the biological scale is under study.
The present work is then a starting point for modeling nanometric bio-machines, such as channels and pumps,
 which control the flux of particles across the cell membrane.

We are indebted to J. Casademunt for fruitful discussions on diffusion  subtleties.
We acknowledge financial help from the Ministerio de Educacion y Ciencia
(Spain) under the  project FIS2006-11452-C03-01 and grant  
FPU-AP-2004-0770 (A. G--M.).


\begin{thebibliography}{1}  

\bibitem{feyn} R.P. Feynman, R.B. Leighton and M. Sands, {\em The Feynman
  Lectures on Physics} (Addison Wesley, Reading, MA, 1963), Vol. {\bf 1}, pp.46.1-46.9.

\bibitem{reim} P. Reimann, Phys. Reports {\bf 361} (2002) p.57-265.

\bibitem{oster} G. Oster,  {\em Darwin's motors}, Nature {\bf 417}, 25 (2002).

\bibitem{liu} D. S. Liu, R. D. Astumian and T. Y. Tsong, 
J. Bio. Chem. {\bf 265}, 7260 (1990).

\bibitem{tsong} T. Y. Tsong and T. D. Xie, 
 Appl. Phys. A{\bf 75}, 345 (2002).

\bibitem{pro} J. Prost, J-F. Chauwin, L. Peliti and A. Ajdari, 
Phys. Rev. Lett. {\bf 72}, 2652 (1994).

\bibitem{astum} R. D. Astumian and I. Derenyi, 
Phys. Rev. Lett. {\bf 86}, 3859 (2001).

\bibitem{astum0} R. D. Astumian, Phys. Rev. Lett. {\bf 91}, 118102 (2003).

\bibitem{kost} I. Kosztin and K. Schulten,
Phys. Rev. Lett. {\bf 93},  238102-1 (2004).


\bibitem{buti1} M. Buttiker, 
Z. Physik B-Comdensed Matter {\bf 68}, 161 (1987).

\bibitem{buti2} Ya M. Blanter and M. Buttiker, 
    Phys. Rev. Lett. {\bf 81}, 4040    (1998).

\bibitem{sbm} A. Gomez-Marin and J. M. Sancho, 
 Phys. Rev. E {\bf 71}, 021101 (2005).


\bibitem{sancho1} J. M. Sancho, M. San Miguel and D. Durr, 
    J. Stat. Phys. {\bf 28}, 291 (1982).



\bibitem{sancho2} J. M. Sancho and A. Sanchez, 
Eur. Phys. J. B. {\bf 16}, 127  (2000).


\bibitem{pavlotis} R. Kupferman, G. A. Pavliotis and A. M. Stuart, 
Phys. Rev. E {\bf 70}, 036120 (2004).


\bibitem{risken}  H. Risken, \emph{The Fokker--Planck equation}, Springer Series in Synergetics, Vol. 18 (Springer, Berlin, 1984).


\bibitem{sims1} B. Nadler, Z. Schuss and A. Singer, 
    Phys. Rev. Lett. {\bf 94}, 218101 (2005).

\bibitem{sims2} R. S. Eisenber, M. M. Klosek and Z. Schuss, 
J. Chem. Phys. {\bf 102}, 1767 (1995).

\bibitem{sims3} P. Szymczak and A. J. C. Ladd, 
    Phys. Rev. E {\bf 69}, 036704 (2004).
    

\bibitem{SPIE} J.M. Sancho and A. Gomez-Marin, 
Proceedings of SPIE, Volume 6602 (2007). 

\end{thebibliography}
\end{document}